# Theoretical analysis on droplet vaporization at elevated temperatures and pressures

Dehai Yu and Zheng Chen[*]

BIC-ESAT, SKLTCS, CAPT, Department of Mechanics and Engineering Science, College of Engineering, Peking University, Beijing 100871, China

**Abstract**

Droplet vaporization has a great impact on combustion of liquid fuels and might greatly affects the engine performance. Physical understanding and theoretical interpretation of the droplet vaporization behavior are decisive to constitute appropriate models for spray combustion. In this work, we proposed a fully transient formulation for the droplet vaporization at wide ranges of temperature and pressure. The governing equations are solved analytically by means of suitable coordinate transformations. The droplet surface temperature and mass fraction of fuel vapor are determined by the matching conditions at the liquid-gas interface. Accordingly, a theoretical model, characterizing the time change of droplet size during its vaporization, is proposed in terms of the unsteadiness factor and droplet vaporization lifetime. The theoretical model successfully reveals the conventional recognition of droplet vaporization characteristics, and its predicted droplet size history agrees well with experimental results reported in the literature. Besides, based on the present theoretical analysis, the experimentally observed dual effect of pressure upon droplet vaporization is analyzed with help of an explicit formula for enthalpy of vaporization indicating its temperature-dependence.

## 1. Introduction

Liquid fuels are widely used in various engines, e.g., diesel engines, jet engines, and liquid rocket engines. To facilitate combustion, the injected fuel is atomized into exceedingly small droplets with diameters in micrometer scale [1-3]. Since chemical reactions occur in gas phase, the fuel droplets should be rapidly evaporated. It has been experimentally observed that only the larger droplets can penetrate the flame zone and burn individually, while the majority of the atomized droplets evaporates in the relatively cool interior of the spray [4]. Therefore, droplet vaporization plays a crucial role in engines and is of primary concern in a variety of multi-phase combustion studies.

---

[*] Corresponding author, Email address: cz@pku.edu.cn, Tel: 86-10-62766232

Experimental studies on droplet vaporization under microgravity conditions without natural convection indicate that the vaporization lifetime decreases monotonically as ambient temperature increases [4, 5]. Moreover, the droplet surface regresses more rapidly at elevated pressures [5, 6]. In experiments, it is very difficult to generate an isolated droplet in stagnant environment without external support, and usually a single droplet is suspended by a thin fiber [5-7]. The suspension technique requires the droplet diameter to be considerably larger than the fiber thickness. Besides, the fiber attachment may cause droplet deformation, which becomes progressively severe with droplet size reducing.

In numerical simulations, different factors that affect droplet vaporization can be isolated and examined independently. Moreover, the evolutions of droplet temperature and diameter can be studied in simulations in wide ranges of pressure and temperature that resemble the engine operating conditions [8-19]. At elevated pressures, various fluid dynamic and thermodynamic non-idealities show profound impacts on the behavior of droplet vaporization [7, 8, 18, 19]. The enthalpy of vaporization decays with the growth of droplet temperature, which results from the increment of ambient pressure [20]. The reduction of enthalpy of vaporization leads to increment of vaporization rate due to the increasing of Spalding transfer number as indicated in $d^2$-law. [8, 21]. Meanwhile, with the increase of pressure, the initial unsteady heating period increases and so does the droplet lifetime [11, 15, 16, 22]. Such a dual effect of the ambient pressure is further affected by the ambient temperature. When the droplet reaches its thermodynamic critical state, the vanishing of liquid-gas interface, the enhanced solvability of environment gas in droplet fluid, and the anomalies in various thermal transport properties may lead to additional intricacy to the vaporization process [8, 16, 18].

In the simulation of spray combustion, simultaneous consideration of a huge number of droplets is a formidable task, which desires theoretical models for droplet vaporization [1, 3, 23]. The primary theory is the $d^2$-law, i.e., the square of droplet diameter decreases linearly with time, which is considered as the leading order solution for a variety of modified theories [19, 24, 25]. By assuming quasi-steady gas-phase process, Law and Sirignano [26-29] conducted a series of theoretical studies on unsteadiness of droplet vaporization and combustion with the emphasis on transient heating effects. They proposed and analyzed two models for droplet heating, i.e., distillation limit and conduction limit. Assuming the density ratio between the gas and liquid, $\rho_g/\rho_l$, as a small parameter, Crespo and Linan [24] conducted asymptotic analysis for droplet vaporization in a stagnant atmosphere considering the unsteady effects in gas phase. They corrected the droplet lifetime by a factor of $\sqrt{\rho_g/\rho_l}$. Zugasti et al. [13] extended the preceding asymptotic analysis to supercritical conditions by discarding the liquid-gas interface and treating the droplet and ambient gas respectively as cold and hot fluid packages. They derived a transcendental equation, which interprets the evolution of the cold package radius and reproduces the $d^2$-law results for isothermal conditions in the cold region.

Nevertheless, at elevated pressures, the prolonged droplet heating stage results in profound unsteadiness in the overall vaporization process. This prevents the direct utilization of $d^2$-law to interpret the droplet size history. Moreover, the density ratio, $\rho_g/\rho_l$, may not be treated as a small parameter at high pressures[30], which reduces the accuracy of the asymptotic analysis. To acquire insightful understanding of droplet vaporization and to accurately model the spray combustion, we need develop an accurate theoretical model which can quantitatively interpret the droplet vaporization at wide ranges of temperature and pressure conditions. The model should explicitly reveal the impacts of various fluid dynamic and thermodynamic properties on the vaporization process. Therefore, it requires a comprehensive theoretical analysis, considering the transient effects and in the meanwhile appropriately dealing with the phase transition at the liquid-gas interface. Besides, the it is recognized that the droplet surface temperature increases with ambient pressure, which, on the one hand, leads to lengthening of droplet heating period, and on the other hand, lowers the enthalpy of vaporization and hence facilitates the vaporization rate. Thereby, the pressure plays a dual effect towards the droplet vaporization. Such analysis is still not in place, which motivates the present study.

The paper is organized as follows. In section 2, the transient formulation for droplet vaporization is solved analytically with the help of appropriate coordinate transformations. The characteristic times for droplet heating and vaporization are derived through the matching conditions. The matching conditions and the Clausius-Clapeyron relation on the droplet surface are solved with the help of those analytical solutions and appropriate evaluation of various thermodynamic and transport properties. In section 3, a theoretical model is proposed, which interprets the time change of droplet size during its vaporization under wide ranges of temperature and pressure. Comparison with experimental results reported in the literature verifies the theoretical model. The duel effect of pressure upon the droplet lifetime is analyzed. The concluding remarks are given in section 4.

## 2. Formulation

We consider a single-component, volatile droplet in a stagnant environment of hot nitrogen with constant and uniform pressure. Since the droplet heating and vaporization take place simultaneously, the formulation for the droplet vaporization process involves mass and energy conservation for both droplet (the liquid phase) and the ambience (the gaseous mixture of fuel vapor and nitrogen) as well as their balance relations across the liquid-gas interface (droplet surface) [20]. The governing equations for gas phase system consist of transport of volatile species, denoted by subscript $F$ and the conservation of energy (in terms of temperature distribution). At the initial instant, no fuel is evaporated and thereby the fuel mass fraction is $Y_F(t = 0, r) = 0$ in the gaseous environment, and both the droplet and the ambience have uniform temperatures, denoted by $T_0$ and $T_\infty$, respectively. As vaporization proceeds, the thermodynamic equilibrium at the evaporating interface is characterized by the Clausius-Clapeyron relation, which correlates the droplet surface temperature, denoted by $T_s = T(r_s, t)$, with the local mass fraction of the evaporated fluid, denoted by $Y_{Fs} = Y_F(r_s, t)$.

### 2.1 Governing equations

For analytical convenience, we assume that the density-weighted mass diffusion coefficients, $\rho D$, thermal conductivities, $\lambda$, and heat capacities at constant pressure of both gas and liquid phases, $c_{pg}$ and $c_{pl}$, are constants. In gas phase, the mass conservation equations for the fuel vapor and inert gas are given by

$$\frac{\partial Y_F}{\partial t} + v \frac{\partial Y_F}{\partial r} = D_g \frac{1}{r^2} \frac{\partial}{\partial r}\left(r^2 \frac{\partial Y_F}{\partial r}\right), \qquad Y_N = 1 - Y_F \qquad (1)$$

where $v$ is the Stefan flow velocity relative to the droplet surface, and $D_g$ is the mass diffusion coefficient of the gaseous mixture. The governing equation for the temperature of the gaseous mixture is

$$\frac{\partial T_g}{\partial t} + v \frac{\partial T_g}{\partial r} = Le_g D_g \frac{1}{r^2} \frac{\partial}{\partial r}\left(r^2 \frac{\partial T_g}{\partial r}\right) \qquad (2)$$

where the Lewis number is defined as $Le_g = \lambda_g/(\rho_g c_{pg} D_g)$.

The governing equation for the temperature of the liquid phase is

$$\frac{\partial T_l}{\partial t} = Le_l D_l \frac{1}{r^2}\frac{\partial}{\partial r}\left(r^2 \frac{\partial T_l}{\partial r}\right) \tag{3}$$

where $D_l$ is the mass diffusion coefficient of the liquid and $Le_l$ is defined in analogy with $Le_g$.

The initial and boundary conditions for the preceding governing equations are specified as

$t = 0$: $\quad\quad\quad\quad Y_F = 0, \quad T_g = T_\infty, \quad T_l = T_0$ (Ic-1)

$r = 0$: $\quad\quad\quad\quad \partial T_l/\partial r = 0$ (Bc-1)

$r = r_s(t)$: $\quad\quad Y_F = Y_{Fs}(t), \quad T_g = T_s(t), \quad T_l = T_s(t)$ (Bc-2)

$r \to \infty$: $\quad\quad\quad\quad Y_F = 0, \quad T_g = T_\infty$ (Bc-3)

where $r_s(t)$ refers to the temporally varying droplet radius.

The governing equations are not in closed form since the temporal variation of surface-related quantities, $Y_{Fs}(t)$, $T_s(t)$, and $r_s(t)$, remain to be determined through the matching conditions at the evaporating interface. The fuel vapor on the droplet surface is partially transported by Stefan flow to downstream and the rest by diffusion to far field. Thereby the mass balance relation is given by

$$\dot{m}_F = \dot{m}_F Y_{Fs} - 4\pi \rho_g D_g r_s^2 \left(\frac{\partial Y_F}{\partial r}\right)_{r=r_s} \tag{4}$$

where the loss of droplet mass is related to the reduction of radius, i.e., $\dot{m}_F = -4\pi \rho_l r_s^2 dr_s/dt$. Meanwhile, energy is transferred from the hot ambience to the droplet surface, partially contributing to the droplet warm-up and the rest providing energy to sustain the vaporization process. Consequently, the energy balance is given by

$$4\pi \lambda r_s^2 \left(\frac{\partial T_g}{\partial r}\right)_{r=r_s} = \dot{m}_F L + 4\pi \lambda_l r_s^2 \left(\frac{\partial T_l}{\partial r}\right)_{r=r_s} \tag{5}$$

where $L$ is the enthalpy of vaporization for the droplet fluid.

We assume that the vaporization occurs at thermodynamically equilibrium state. Therefore, on the droplet surface, the relationship between mass fraction of fuel vapor and temperature is given by the Clausius-Clapeyron relation, i.e.,

$$\frac{Y_{Fs}/W_F}{Y_{Fs}/W_F + (1-Y_{Fs})/W_N} = \frac{p_n}{p}\exp\left[\frac{L(T_s)}{R}\left(\frac{1}{T_{b,n}} - \frac{1}{T_s}\right)\right] \tag{6}$$

where $W$ is the molar mass of each species, $T_{bn}$ the boiling point temperature of the droplet fluid under the normal pressure $p_n$. Supplementing the matching conditions and phase equilibrium relation to the governing equations, the mathematical formulation of the droplet vaporization problem is in closed form, and hence the solution procedure could be initiated.

**2.2 Solutions to the gas phase**

The evaporation results in the reduction of droplet radius. For mathematical convenience in dealing with the moving boundary problem, we introduce a scaled coordinate in analogy to that adopted by Law and Sirignano [27], i.e.,

$$\sigma = \frac{r}{r_s(t)}, \qquad \tau = \frac{D_g t}{r_s^2} \tag{7}$$

The characteristic time scale, $r_s^2/D_g$, reveals the gas-phase unsteadiness during the droplet vaporization process. Applying the coordinate transformation (7) to Eqs. (1) and (2), we have

$$\frac{\partial Y_F}{\partial \tau} - \frac{1}{2}\left(\frac{1}{\sigma^2}\frac{\rho_l}{\rho_g} + \sigma\right)\frac{1}{D_g}\frac{dr_s^2}{dt}\frac{\partial Y_F}{\partial \sigma} = \frac{1}{\sigma^2}\frac{\partial}{\partial \sigma}\left(\sigma^2 \frac{\partial Y_F}{\partial \sigma}\right) \tag{8}$$

$$\frac{\partial T_g}{\partial \tau} - \frac{1}{2}\left(\frac{1}{\sigma^2}\frac{\rho_l}{\rho_g} + \sigma\right)\frac{1}{D_g}\frac{dr_s^2}{dt}\frac{\partial T_g}{\partial \sigma} = Le_g \frac{1}{\sigma^2}\frac{\partial}{\partial \sigma}\left(\sigma^2 \frac{\partial T_g}{\partial \sigma}\right) \tag{9}$$

As the radial coordinate $r$ is replaced by the normalized coordinate $\sigma$, the moving boundary effect is transformed to an equivalent convection term. The initial and boundary conditions for Eqs. (8) and (9) are specied as

| | | | |
|---|---|---|---|
| $\tau = 0$ | $Y_F = 0,$ | $T_g = T_\infty$ | (Ic-1') |
| $\sigma = 1$ | $Y_F = Y_{Fs}(t),$ | $T_g = T_s(t)$ | (Bc-2') |
| $\sigma \to \infty$ | $Y_F = 0,$ | $T_g = T_\infty$ | (Bc-3') |

The first order derivatives impose substantial difficulties in delaing with Eqs. (8) and (9) analytically. Whereas, such mathematical issue could be alleviated by introducing the following coordinate transform

$$\xi_g = \left[\int_1^\sigma F_g(\sigma')d\sigma'\right] \times \left[\int_1^\infty F_g(\sigma')d\sigma'\right]^{-1} \tag{10}$$

where

$$F_g(\sigma) = \exp\left\{-\int_1^\sigma \left[\frac{2}{\sigma'} + \frac{1}{2}\left(\frac{1}{\sigma'^2}\frac{\rho_l}{\rho} + \sigma'\right)\frac{1}{D_g}\frac{dr_s^2(t)}{dt}\right]d\sigma'\right\} \tag{11}$$

Applying the coordination transformation (10) to Eqs. (8) and (9), we have

$$\frac{\partial Y_F}{\partial \tau} = D_{\text{eff}} \frac{\partial^2 Y_F}{\partial \xi_g^2}, \qquad \frac{\partial T_g}{\partial \tau} = Le_g D_{\text{eff}} \frac{\partial^2 T_g}{\partial \xi_g^2} \qquad (12)$$

where the effective diffusion coefficient is defined as

$$D_{\text{eff}} = \left\{ F_g(\sigma) \left[ \int_1^\infty F_g(\sigma') d\sigma' \right]^{-1} \right\}^2 \qquad (13)$$

It is a function of $\sigma$ and thus varies with $\xi_g$. On the droplet surface where evaporation occurs, the solutions for $Y_F$ and $T$ are to be substituted into the matching conditions. This implies that the accuracy of the solutions for $Y_F$ and $T$ remote from the droplet surface has insignificant impact upon determination of the droplet vaporization rate, which relies upon the analysis at evaporating interface. It allows us to choose a representative value for $D_{\text{eff}}$ that characterizes its contirbution to $Y_{FS}$ and $T_s$ evaluated by their analytical solutions to Eqs. (8) and (9). Accordingly, we may specify the characteristic effective diffusion coefficient on the droplet surface where $\sigma = 1$, giving

$$D_{\text{eff}}^c = \left[ \int_1^\infty F(\sigma') d\sigma' \right]^{-2} \qquad (14)$$

The boundary conditions are revised as

$\xi_g = 0:$ $\qquad Y_F = Y_{Fs}(\tau), \qquad T_g = T_s(\tau)$ (Bc-2")

$\xi_g = 1:$ $\qquad Y_F = 0, \qquad T_g = T_\infty$ (Bc-3")

With the help of $D_{\text{eff}}$, the Eq. (12) could be solved analytcally subject to the initial condition (Ic-1') and boundary conditions (Bc-2") and (Bc-3"), giving

$$Y_F(\xi_g, \tau) = Y_{Fs}(\tau) - \xi_g Y_{Fs}(\tau) + 2 \sum_{n=1}^\infty \sin(n\pi x) \, e^{-D_{\text{eff}}^c n^2 \pi^2 \tau} R_n^Y(\tau) \qquad (15)$$

$$T_g(\xi_g, \tau) = T_s(\tau) + \xi_g [T_\infty - T_s(\tau)] + 2 \sum_{n=1}^\infty \sin(n\pi x) \, e^{-Le_g D_{\text{eff}}^c n^2 \pi^2 \tau} R_n^T(\tau) \qquad (16)$$

where $R_n^Y(\tau)$ and $R_n^T(\tau)$ are

$$R_n^Y(\tau) = -\frac{1}{n\pi} \left( \frac{\partial Y_{Fs}}{\partial \tau} \frac{e^{D_{\text{eff}}^c n^2 \pi^2 \tau} - 1}{D_{\text{eff}}^c n^2 \pi^2} - \int_0^t \frac{e^{D_{\text{eff}}^c n^2 \pi^2 \tau} - 1}{D_{\text{eff}}^c n^2 \pi^2} \frac{\partial^2 Y_{Fs}}{\partial \tau^2} d\tau \right) \qquad (17)$$

$$R_n^T(\tau) = \int_0^1 T_\infty \sin(n\pi\xi_g)\, d\xi_g - \frac{1}{n\pi}[T_s(0) - (-1)^n T_\infty]$$

$$-\frac{1}{n\pi}\left(\frac{\partial T_s}{\partial \tau}\frac{e^{Le_g D_{\text{eff}}^c n^2\pi^2\tau} - 1}{Le_g D_{\text{eff}}^c n^2\pi^2} - \int_0^t \frac{e^{Le_g D_{\text{eff}}^c n^2\pi^2\tau} - 1}{Le_g D_{\text{eff}}^c n^2\pi^2}\frac{\partial^2 T_s}{\partial \tau^2}d\tau\right) \quad (18)$$

On the droplet surface, the fuel vapor is transported to the ambience by convection and diffusion (loss contribution) and meanwhile supplied by continuous evaporation from liquid droplet (gain contribution). Therefore, we argue that dynamic equilibrium is maintained on the droplet surface. It implies that except at a very brief period, during which the phase equilibrium condition is established, the time change of $Y_{Fs}$ tends to be inconsiderable than the unsteadiness resulting from the droplet heating in vaporization process [28, 29]. The same argument applies to the droplet surface temperature $T_s$ according to the Clausius-Clapeyron relation. Accordingly, and we can neglect the terms containing the second order derivatives of $Y_{Fs}$ and $T_s$ in Eqs. (17) and (18). Substituting those simplifications into Eqs. (15) and (16) yields the following solutions for $Y_F$ and $T_g$

$$Y_F(\xi_g, \tau) = Y_{Fs}(\tau) - \xi_g Y_{Fs}(\tau) - \frac{2}{D_{\text{eff}}^c}\frac{\partial Y_{Fs}}{\partial \tau}\sum_{n=1}^{\infty}\frac{\sin(n\pi\xi_g)}{n^3\pi^3}$$

$$+ 2\sum_{n=1}^{\infty}\frac{\sin(n\pi\xi_g)\, e^{-D_{\text{eff}}^c n^2\pi^2\tau}}{n\pi}\left(\frac{1}{n^2\pi^2 D_{\text{eff}}^c}\frac{\partial Y_{Fs}}{\partial \tau}\right) \quad (19)$$

$$T_g(\xi_g, \tau) = T_s(\tau) + \xi_g[T_\infty - T_s(\tau)] - \frac{2}{Le_g D_{\text{eff}}^c}\frac{\partial T_s}{\partial \tau}\sum_{n=1}^{\infty}\frac{\sin(n\pi\xi_g)}{n^3\pi^3}$$

$$+ 2\sum_{n=1}^{\infty}\frac{\sin(n\pi\xi_g)}{n\pi}e^{-Le_g D_{\text{eff}}^c n^2\pi^2\tau}\left[T_\infty - T_s(0) + \frac{1}{n^2\pi^2 Le_g D_{\text{eff}}^c}\frac{\partial T_s}{\partial \tau}\right] \quad (20)$$

On the droplet surface where $\sigma = 1$, the derivatives of $Y_F$ and $T_g$ with respect to $\sigma$ shall be calculated through chain's rule, i.e.,

$$\left(\frac{\partial Y_F}{\partial \sigma}\right)_{\sigma=1} = \left(\frac{\partial Y_F}{\partial \xi_g}\right)_{\xi=0}\left(\frac{d\xi_g}{d\sigma}\right)_{\sigma=1}, \quad \left(\frac{\partial T}{\partial \sigma}\right)_{\sigma=1} = \left(\frac{\partial T}{\partial \xi_g}\right)_{\xi=0}\left(\frac{d\xi_g}{d\sigma}\right)_{\sigma=1} \quad (21)$$

With the help of Eq. (10) and evaluating the derivatives of Eqs. (19) and (20), we have

$$\left(\frac{\partial Y_F}{\partial \sigma}\right)_{\sigma=1} = -\left[Y_{Fs}(\tau) + \frac{1}{3}\frac{1}{D_{\text{eff}}^c}\frac{\partial Y_{Fs}}{\partial \tau}\left(1 - e^{-D_{\text{eff}}^c \pi^2\tau}\right)\right]\left[\int_1^\infty F_g(\sigma)d\sigma\right]^{-1} \quad (22)$$

$$\left(\frac{\partial T_g}{\partial \sigma}\right)_{\sigma=1} = \left\{T_\infty - T_s(\tau) - \frac{1}{3}\frac{1}{Le_g D_{\text{eff}}^c}\frac{\partial T_s}{\partial \tau}\left(1 - e^{-Le_g D_{\text{eff}}^c \pi^2 \tau}\right)\right.$$
$$\left. + [T_\infty - T_s(0)]\left[\vartheta\left(3, 0, e^{-Le_g D_{\text{eff}}^c \pi^2 \tau}\right) - 1\right]\right\}\left[\int_1^\infty F_g(\sigma)d\sigma\right]^{-1} \quad (23)$$

where $\vartheta$ refers to the elliptic theta function. Since the quasi-steady state in gas phase establishes swiftly during the droplet vaporization, we make the subsequent approximation

$$\sum_{n=1}^\infty \frac{e^{-Le_g D_{\text{eff}}^c n^2 \pi^2 \tau}}{n^2 \pi^2} \to e^{-Le_g D_{\text{eff}}^c \pi^2 \tau}\sum_{n=1}^\infty \frac{1}{n^2 \pi^2} = \frac{1}{6}e^{-Le_g D_{\text{eff}}^c \pi^2 \tau} \quad (24)$$

during derivation of Eqs. (22) and (23).

Known the integral of $F_g(\sigma)$ from 1 to $\sigma = \infty$, the characteristic value of effective diffusion coefficient can be calculated, and subsequently the determination of $(\partial Y_F/\partial \sigma)_{\sigma=1}$ and $(\partial T_g/\partial \sigma)_{\sigma=1}$ according to Eqs. (22) and (23). Analyzing Eq. (11), it is noted that the linear term $(\sigma/D_g)(dr_s^2/dt)$ actually results from the coordinate transformation from the $(r,t)$- to $(\sigma,\tau)$-space. When approaching to the end of the droplet vaporization, i.e., $r_s \to 0$, the coordinate $\sigma$, according to Eq. (7), becomes exceedingly large, even for moderate radial distance $r$. Such pure mathematical effect leads to the divergence of the $F_g(\sigma)$ at large values of $\sigma$, and has no contribution to the physical interpretation of the droplet vaporization process. Thereby, we remove this linear term in the estimation of the integral of $F_g(\sigma)$, yielding

$$\int_1^\infty F(\sigma)d\sigma \approx \int_1^\infty \exp\left\{-\int_1^\sigma \left[\frac{2}{\sigma'} + \frac{1}{2}\frac{1}{\sigma'^2}\frac{\rho_l}{\rho_g}\frac{1}{D_g}\frac{dr_s^2(t)}{dt}\right]d\sigma'\right\}d\sigma$$
$$= \frac{1}{a_g}(e^{a_g} - 1) \quad (25)$$

where $a_g$ is

$$a_g = -\frac{1}{2}\frac{1}{D_g}\frac{dr_s^2}{dt}\frac{\rho_l}{\rho_g} > 0 \quad (26)$$

Substituting Eq. (25) into Eqs. (22) and (23), we obtain

$$\left(\frac{\partial Y_F}{\partial \sigma}\right)_{\sigma=1} = -\left[Y_{Fs}(\tau) + \frac{1}{3}\frac{1}{D_{\text{eff}}^c}\frac{\partial Y_{Fs}}{\partial \tau}\left(1 - e^{-D_{\text{eff}}^c \pi^2 \tau}\right)\right]\frac{a_g}{e^{a_g} - 1} \quad (27)$$

$$\left(\frac{\partial T_g}{\partial \sigma}\right)_{\sigma=1} = \left\{T_\infty - T_s(\tau) - \frac{1}{3}\frac{1}{Le_g D_{\text{eff}}^C}\frac{\partial T_s}{\partial \tau}\left(1 - e^{-Le_g D_{\text{eff}}^C \pi^2 \tau}\right)\right.$$
$$\left. + [T_\infty - T_s(0)]\left[\vartheta\left(3, 0, e^{-Le_g D_{\text{eff}}^C \pi^2 \tau}\right) - 1\right]\right\}\frac{a_g}{e^{a_g} - 1} \quad (28)$$

Equations (27) and (28) are to be substituted into the mass and energy balance relations on the droplet surface, respectively.

## 2.3 Solutions to the liquid phase

By defining $T = rT_l$, Eq. (3) can be written in the form of one-dimensional heat conduction in Cartesian coordinate, i.e.,

$$\frac{\partial T}{\partial t} = D_l Le_l \frac{\partial^2 T}{\partial r^2} \quad (29)$$

Applying the coordinate transform (7) to Eq. (29), we have

$$\frac{\partial T}{\partial \tau} - \frac{1}{2}\frac{\sigma}{D_g}\frac{dr_s^2(t)}{dt}\frac{\partial T}{\partial \sigma} = \frac{Le_l D_l}{D_g}\frac{\partial^2 T}{\partial \sigma^2} \quad (30)$$

In analogy to Eq. (10), we define a new coordinate $\xi_l$ as

$$\xi_l = \frac{\text{erfi}(\sqrt{a_l}\sigma)}{\text{erfi}(\sqrt{a_l})} \quad (31)$$

where "erfi" refers to the imaginary error function, and the factor $a_l$ is defined as

$$a_l = -\frac{1}{4}\frac{1}{D_l Le_l}\frac{dr_s^2}{dt} \quad (32)$$

Applying the coordinate transform (31) to Eq. (30), the first order derivative no longer appears, giving

$$\frac{\partial T}{\partial \tau} = \lambda_{l,\text{eff}}\frac{d^2 T}{d\xi_l^2} \quad (33)$$

where the effective heat conduction coefficient denotes for

$$\lambda_{l,\text{eff}} = \left[\frac{1}{e^{-a_l \sigma^2}}\frac{\sqrt{\pi}\text{erfi}(\sqrt{a_l})}{2\sqrt{a_l}}\right]^{-2}\frac{D_l Le_l}{D_g} \quad (34)$$

Following the same procedure as we derive $D_{\text{eff}}^C$, the characteristic value for effective heat conduction coefficient, denoted by $\lambda_{l,\text{eff}}^c$, can be specified by setting $\sigma = 1$. The initial and boundary conditions for Eq. (33) are

$$\tau = 0 \qquad\qquad T = \sigma(\xi_l) r_0 T_0 \qquad\qquad \text{(Ic-i)}$$

$$\xi_l = 0 \qquad\qquad T = 0 \qquad\qquad \text{(Bc-ii)}$$

$$\xi_l = 1 \qquad\qquad T = r_s(\tau) T_s(t) \qquad\qquad \text{(Bc-iii)}$$

where $r_0$ is the initial radius of the droplet.

The functional relationship between $\sigma$ and $\xi_l$ shall be obtained by inversely solving Eq. (31), formally yielding

$$\sigma = -\frac{i\,\text{erf}^{-1}\left[i\xi_l \text{erfi}\left(\sqrt{a_l}\right)\right]}{\sqrt{a_l}} \qquad (35)$$

where $\text{erf}^{-1}$ denotes the inverse function of error function. Equation (32) shows that $a_l$ involves the surface regression rate of the evaporating droplet, $dr_s^2/dt$, which can be considered as a small quantity during droplet vaporization. This implies that the right-hand side of Eq. (35) can be approximated by power series in terms of $a_l$, giving

$$\sigma = \xi_l + \frac{1}{3}(\xi_l - \xi_l^3) a_l + O(a_l^2) \qquad (36)$$

The initial condition (Ic-i) is accordingly modified to

$$\tau = 0 \qquad\qquad T = \left[\xi_l + \frac{a_l}{3}(\xi_l - \xi_l^3)\right] r_0 T_0 \qquad\qquad \text{(Ic-i')}$$

which we shall adopt in the subsequent solution of the liquid phase system.

Equating the effective heat conduction coefficient to its characteristic value, Eq. (33) can be solved analytically subject to the initial condition (Ic-i') and boundary conditions (Bc-ii) and (Bc-iii), giving

$$T(\xi_l, \tau) = \left[\xi_l + \frac{a_l}{3}(\xi_l - \xi_l^3)\right] r_s(\tau) T_s(\tau) + 2\sum_{n=1}^{\infty} \sin(n\pi\xi_l) R_{n,l}(\tau)\, e^{-\lambda_{l,\text{eff}}^c n^2 \pi^2 \tau} \qquad (37)$$

where, considering the dynamic equilibrium on the droplet surface, the quantity $R_{n,l}$ can be written as

$$R_{n,l}(\tau) = \frac{2 a_l (-1)^{n+1}}{n^3 \pi^3} T_0 r_0 + (-1)^n \frac{e^{\lambda_{l,\text{eff}}^c n^2 \pi^2 \tau} - 1}{\lambda_{l,\text{eff}}^c n^3 \pi^3} \frac{d}{d\tau}[r_s(\tau) T_s(\tau)] \qquad (38)$$

Substituting Eq. (38) into Eq. (37) and evaluating the derivative $dT/d\xi_l$ on droplet surface where $\xi_l = 1$, we have

$$\left(\frac{\partial T}{\partial \xi_l}\right)_{\xi_l=1} \approx \left(1 - \frac{2}{3}a_l\right) r_s T_s - \frac{1}{3} T_0 r_0 a_l e^{-\lambda_{l,\text{eff}}^c \pi^2 \tau}$$
$$+ \frac{1}{3} \frac{1}{\lambda_{l,\text{eff}}^c} \left(r_s \frac{dT_s}{d\tau} + T_s \frac{dr_s}{d\tau}\right)\left(1 - e^{-\lambda_{l,\text{eff}}^c \pi^2 \tau}\right) \tag{39}$$

where the approximation (24) is adopted when dealing with the summation terms. Using chain's rule of differentiation, the derivative of $T_l$ with respect to $\sigma$ can be evaluated by

$$\left(\frac{\partial T_l}{\partial \sigma}\right)_{\sigma=1} = \frac{1}{r_s \sigma} \left(\frac{\partial T}{\partial \xi_l}\right)_{\xi_l=1} \left(\frac{\partial \xi_l}{\partial \sigma}\right)_{\sigma=1} - \frac{T}{r_s \sigma^2}$$
$$= \left[\left(1 - \frac{2}{3}a_l\right) T_s - \frac{1}{3} T_0 \frac{r_0}{r_s} a_l e^{-\lambda_{l,\text{eff}}^c \pi^2 \tau}\right. \tag{40}$$
$$\left. + \frac{1}{3} \frac{1}{\lambda_{l,\text{eff}}^c} \left(\frac{dT_s}{d\tau} + \frac{1}{2} \frac{T_s}{D_g} \frac{dr_s^2}{dt}\right)\left(1 - e^{-\lambda_{l,\text{eff}}^c \pi^2 \tau}\right)\right] \frac{2\sqrt{a_l} e^{-a_l}}{\sqrt{\pi} \text{erfi}(\sqrt{a_l})} - T_s$$

**2.4 Analysis on the droplet surface**

The mass balance relation on the interface in the scaled coordinate $(\tau, \sigma)$ is given by

$$\frac{\rho_l}{\rho_g} \frac{1}{D_g} \frac{dr_s^2}{dt} (1 - Y_{Fs}) = 2 \left(\frac{\partial Y_F}{\partial \sigma}\right)_{\sigma=1} \tag{41}$$

After a swift induction period in the order of $O(1/D_{\text{eff}}^c \pi^2)$, the gas phase unsteadiness decays to be negligible. Accordingly, we can remove the exponentially decaying term in the coefficient of $dY_{Fs}/dt$ in Eq. (27), giving,

$$\left(\frac{\partial Y_F}{\partial \sigma}\right)_{\sigma=1} = -\left(Y_{Fs} + \frac{1}{3} \frac{1}{D_{\text{eff}}^c} \frac{\partial Y_{Fs}}{\partial \tau}\right) \frac{a_g}{e^{a_g} - 1} \tag{42}$$

Substituting the above expression into Eq. (41), we obtain an ordinary differential equation for the temporal variation of $Y_{Fs}$, i.e.,

$$\frac{\partial Y_{Fs}}{\partial \tau} = -3 D_{\text{eff}}^c (e^{a_g} Y_{Fs} + 1 - e^{a_g}) \tag{43}$$

The droplet size changes slightly during the initial build-up of fuel vapor and the elevation of droplet surface temperature, which implies that the quantity $a_g$ might be considered as a constant. Solving Eq. (43) subject to the initial condition of $Y_{Fs}(0) = 0$ yields

$$Y_{Fs} = (1 - e^{-a_g})\left(1 - e^{-3 e^{a_g} D_{\text{eff}}^c \tau}\right) \tag{44}$$

It is seen that after an induction period characterized by $1/3e^{a_g}D^c_{\text{eff}}$, Eq. (44) asymptotes to

$$Y^0_{FS} = 1 - e^{-a_g} \tag{45}$$

which shall be considered as the mass fraction of fuel vapor when the droplet vaporization occurring at quasi-steady state. Substituting the asymptotic value of $Y^0_{FS}$ into the Clausius-Clapeyron relation, given by Eq. (6), the asymptotic temperature on the droplet surface can be obtained as

$$T^0_s = T_{bn}\left(1 + \frac{RT_{bn}}{L(T^0_s)}\left\{\ln\frac{p_n}{p} + \ln\left[1 + \frac{W_F}{W_N(e^{a_g} - 1)}\right]\right\}\right)^{-1} \tag{46}$$

For droplet evaporating at high temperature and pressure conditions, the quasi-steady vaporization is preceded by a heating stage, during which the droplet temperature is progressively increased to a saturated value, equivalently to $T^0_s$ given by Eq. (46). During the heating stage, vaporization rate tends to be indiscernible compared with that in the subsequent stage of quasi-steady vaporization. Consequently, the duration for increasing the droplet surface temperature would be solved by modifying the energy balance relation, given by Eq. (5), where the vaporization effect is removed, yielding

$$\lambda_g\left(\frac{\partial T_g}{\partial \sigma}\right)_{\sigma=1} = \lambda_l\left(\frac{\partial T_l}{\partial \sigma}\right)_{\sigma=1} \tag{47}$$

Since vaporization is negligible, it implies that $a_g \approx 0$ and $a_l \approx 0$. Substituting the temperature gradients on each side of the droplet surface into Eq. (47), we obtain the following ordinary differential equation for $T_s$:

$$\frac{dT_s}{d\tau} = \frac{3(\lambda_g/D_g)(T_\infty - T_s)}{\left(1 - e^{-\lambda_l\pi^2\tau/\rho_l c_{vl} D_g}\right)\rho_l c_{vl} + \rho_g c_{pg}} \tag{48}$$

Solving for $T_s$ subject to the initial condition $T_s(0) = T_0$ and considering the duration when $T_s$ is increased from $T_0$ to $T^0_s$, it gives a partial heating time

$$t^{(1)}_{\text{heating}} = \frac{r^2_0 \rho_g c_{pg}}{\lambda_l \pi^2}\left(\frac{T_\infty - T_0}{T_\infty - T^0_s}\right)^{\lambda_l\pi^2/3\lambda_g} \tag{49}$$

We have utilized the simplification that $\rho_g c_{pg}/\rho_l c_{vl} \ll 1$, which is generally valid unless the vaporization takes place at the near-critical condition [30, 31].

During the quasi-steady vaporization, the droplet heating effect tends to be inconsiderable, and hence the droplet temperature is approximately uniform. Thereby, the rest duration of the heating stage would be estimated through

$$\frac{4\pi r_0^3}{3} c_{vl}(T_s^0 - T_0) = \int_0^{t_{\text{heating}}^{(2)}} 4\pi r_0^2 \lambda_g \left(\frac{\partial T_g}{\partial \sigma}\right)_{\sigma=1} dt \tag{50}$$

Solving Eq. (50) for $t_{\text{heating}}^{(2)}$, we obtain

$$t_{\text{heating}}^{(2)} = \frac{c_{vl}\rho_l r_0^2}{3\lambda_g} \frac{T_s^0 - T_0}{T_\infty - T_s^0} \tag{51}$$

Consequently, the total time lapse of the heating stage would be

$$t_{\text{heating}} = t_{\text{heating}}^{(1)} + t_{\text{heating}}^{(2)}$$
$$= \frac{c_{vl}\rho_l r_0^2}{3\lambda_g} \left[\frac{T_s^0 - T_0}{T_\infty - T_s^0} + 3\frac{\rho_g c_{pg}}{\rho_l c_{vl}} \frac{\lambda_g}{\lambda_l} \left(\frac{T_\infty - T_0}{T_\infty - T_s^0}\right)^{\lambda_l \pi^2 / 3\lambda_g}\right] \tag{52}$$

In the quasi-steady vaporization stage, the droplet temperature almost remains constant, implying that heat transfer from the hot ambience is mainly utilized to evaporate the droplet. Therefore, the energy balance relation can be revised from Eq. (5) by removing the heating effect, giving

$$\left(\frac{\partial T_g}{\partial \sigma}\right)_{\sigma=1} = -\frac{1}{2}\frac{\rho_l}{\lambda_g}\frac{dr_s^2(t)}{dt}L(T_s) \tag{53}$$

Substituting $(\partial T_g/\partial \sigma)_{\sigma=1}$ from Eq. (28) into Eq. (53), we obtain an ordinary differential equation for $r_s^2$, which is analogous to the classic $d^2$-law characterizing the droplet vaporization, i.e.,

$$\frac{dr_s^2}{dt} = -\frac{2\lambda_g}{\rho_l c_{pg}} \frac{1}{Le_g} \ln\left[1 + \frac{Le_g c_{pg}(T_\infty - T_s^0)}{L(T_s^0)}\right] \tag{54}$$

Solving for time duration of quasi-steady vaporization, $t_{\text{vaporization}}$, we obtain

$$t_{\text{vaporization}} = \frac{1}{2}\frac{\rho_l c_{pg} Le_g r_0^2}{\lambda_g \ln[1 + Le_g c_{pg}(T_\infty - T_s^0)/L(T_s^0)]} \tag{55}$$

In terms of Eqs. (52) and (55), the overall droplet lifetime can be evaluated in the following explicit form:

$$t_{\text{total}} = t_{\text{heating}} + t_{\text{vaporization}}$$
$$= \frac{\rho_l}{\rho_g}\frac{r_0^2}{D_g}\left\{\frac{1}{3Le_g}\frac{c_{vl}}{c_{pg}}\frac{T_s^0 - T_0}{T_\infty - T_s^0} + \frac{\rho_g c_{pg}}{\rho_l c_{vl}}\frac{\lambda_g}{\lambda_l}\left(\frac{T_\infty - T_0}{T_\infty - T_s^0}\right)^{\lambda_l \pi^2 / 3\lambda_g}\right. \tag{56}$$
$$\left. + \frac{1}{2}\frac{1}{\ln[1 + Le_g c_{pg}(T_\infty - T_s^0)/L(T_s^0)]}\right\}$$

Acquiring the surface regression rate, given by Eq. (54), the asymptotic droplet surface temperature, according to Eq. (46), can be written in the subsequent form

$$T_s^0 = T_{bn}\left(1 + \frac{RT_{bn}}{L(T_s^0)}\left\{\ln\frac{p_n}{p} + \ln\left[1 + \frac{W_F L(T_s^0)}{W_N Le_g c_{pg}(T_\infty - T_s^0)}\right]\right\}\right)^{-1} \quad (57)$$

Therefore, $T_s^0$ shall be determined iteratively through Eq. (57) when the temperature dependence of enthalpy of vaporization is known.

## 2.5 Thermodynamic and transport properties

The enthalpy of vaporization can be directly solved through

$$L = \int_{V_m^l}^{V_m^v}\left[T\left(\frac{\partial P}{\partial T}\right)_{V_m} - P\right]dV_m + P_e(V_m^v - V_m^l) \quad (58)$$

where $V_m^l$ and $V_m^v$ denote the molar volumes of liquid and gas phase, respectively, and $P_e$ is the equilibrium vapor pressure. Those equilibrium states can be iteratively calculated by simultaneously solving an equation of state that is accurate at both liquid and gas phase, e.g., Soave-Redich-Kwong or Peng-Robinson equation of states. It is acknowledged that Eq. (58) cannot interpret the temperature-dependence of enthalpy of vaporization in a mathematically explicit form.

Fortunately, the enthalpy of vaporization can also be evaluated through an analytical formula [32], given by

$$\frac{L}{L_0} = \left[1 - f_E\left(\frac{q_{l,R}}{2} - 1\right)t\right]^{1-t}\left[f_E\frac{q_{l,R}}{2(1-\alpha)}(1-t)^{1-\alpha} + \left(1 - \frac{1}{2}f_E q_{l,R}\right)(1-t)^\beta\right]^t \quad (59)$$

where $L_0$ is the enthalpy of vaporization at reference temperature $T_R$. The $\alpha$ and $\beta$ are the critical exponents characterizing the heat capacity at constant volume and the density difference when the fluid is close to its thermodynamic critical state, i.e., $c_v \sim (T_c - T)^{-\alpha}$ and $\rho_l - \rho_c \sim (T_c - T)^\beta$. The literature on the critical phenomena suggests that the approximate values of $\alpha$ of $\beta$ for various fluids are 0.1096 and 0.3265, respectively [33]. According to the X-ray scattering experimental studies on liquid configurations, the coordination number, denoted by $q$, changes in a restricted range between 9 and 11 for most fluids [34]. Since the reference state is remote from the critical point, reference coordination number, $q_{l,R}$, tends to be greater in magnitude, which leads us to specify $q_{l,R} = 11$ in the subsequent discussion. The energy factor $f_E$ and normalized temperature $t$ are defined in terms of reference and critical temperatures

$$f_E = \frac{R(T_c - T_R)}{L_0}, \qquad t = \frac{T - T_R}{T_c - T_R} \tag{60}$$

Therefore, with the knowledge of both reference state and critical temperature, i.e., $L_0$ and $T_c$, the enthalpy of vaporization can be calculated through Eq. (59) at the surface temperature $T_s^0$.

Since the gas-phase system includes both fuel vapor and nitrogen, the heat capacity at constant pressure must consider the effect of mixture, i.e.,

$$c_{pg} = Y_{Fs}^0 c_{pg,F} + (1 - Y_{Fs}^0) c_{pg,N} \tag{61}$$

where $c_{pg,F}$ and $c_{pg,N}$ denote the heat capacities at constant pressure for pure fuel vapor and inert gas, respectively. The heat capacity varies with temperature, and thus a characteristic temperature is required to determine $c_{pg,F}$ and $c_{pg,N}$. Since the ambient temperature is considerably higher than the surface temperature, which must be less than the boiling temperature at the current pressure, it is necessary to consider the enhancement effect due to the high temperature environment. For a practical estimation, we adopt the conventional 1/3 rule to evaluate the characteristic temperature that is utilized to determine various transport properties, denoted by $q$ in general, i.e.,

$$q = q(T_{ch}), \qquad T_{ch} = \frac{2}{3} T_s^0 + \frac{1}{3} T_\infty \tag{62}$$

The models that interpret the thermodynamic and transport properties as functions of temperature are given in ref [16].

According to Eq. (61), the heat capacity depends upon the mass fraction of evaporated fuel vapor. This implies that the determination of $T_s^0$ and $Y_{Fs}^0$ are inherently coupled and hence shall be calculated simultaneously and iteratively. With the knowledge of $T_s^0$ and ambient pressure $p$, the mass densities of droplet and the gaseous mixture can be determined with the help of an appropriate equation of state that is accurate over the whole density range of the concerned fluid. A possible candidate is the Peng-Robinson equation of state, whose mathematical form is given by

$$p = \frac{RT}{V_m - b} - \frac{a}{V_m(V_m + b) + b(V_m - b)} \left[1 + f_\omega \left(1 - T_r^{1/2}\right)\right]^2 \tag{63}$$

where the parameters are determined by the critical properties and acentric factor $\omega$

$$\begin{aligned} f_\omega &= 0.37464 + 1.54226\omega + 0.26992\omega^2 \\ a &= 0.45724 \frac{R^2 T_c^2}{p_c^2} \\ b &= 0.07780 \frac{R T_c}{p_c} \end{aligned} \tag{64}$$

For situations with $T < T_c$ and $p < p_c$, solving Eq. (63) gives three values of molar densities

$(1/V_m)$, which can be transformed into mass-based densities. The density of gas-phase mixture depends upon the mass fraction of fuel vapor through

$$\rho_g = Y_{Fs}^0 \rho_{g,F} + (1 - Y_{Fs}^0)\rho_{g,N} \qquad (65)$$

Substituting those specified thermodynamic and transport properties into Eqs. (52), (55), and (56), the characteristic time scales of droplet vaporization, i.e., $t_{\text{heating}}$, $t_{\text{vaporization}}$ and $t_{\text{total}}$, can be readily calculated.

## 3 Results and Discussion

To verify the present theoretical analysis, we specify our conditions identical to those in Nomura et al.'s experiments [6], which considered an n-heptane droplet vaporizing at nitrogen environment under microgravity condition. In experiments, the droplet was suspended by a thin fiber, thus the overall droplet vaporization lifetime must be extrapolated from the droplet size history to its vanishing point. In Fig. 1, the scaled vaporization lifetimes (divided by the square of the initial droplet diameter), calculated by Eq. (56) at various temperatures and pressures, are compared with those extrapolated from experimental data. It is seen that the theoretical model can accurately predict the droplet lifetime in wide ranges of temperature and pressure. Both theoretical model and experimental results indicate that the droplet lifetime progressively decay as increasing the ambient temperature. Whereas, at relative low temperatures, the droplet lifetime would be significantly prolonged when the ambient pressure is elevated.

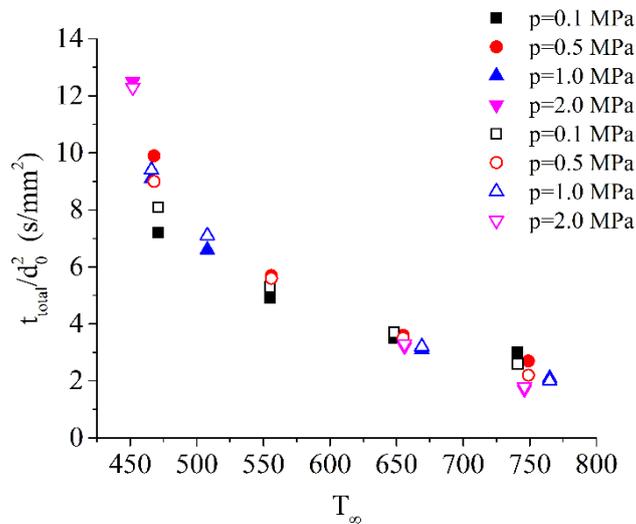

Figure 1. Comparison of droplet lifetime calculated by theoretical model (denoted by hollow symbols)

and those extrapolated from experimental data (denoted by solid symbols) in [6].

The increase of ambient pressure leads to growth of the saturated temperature on the droplet surface, $T_s^0$, implying that more heat is required to warm up the droplet before its arrival at the quasi-steady vaporization state. Accordingly, it increases the heating time and the unsteadiness of the vaporization process. Nomura et al. [6] defined an unsteadiness factor as the ratio of droplet heating time to its lifetime, i.e.,

$$\Phi_i = \frac{t_{\text{heating}}}{t_{\text{total}}} \tag{66}$$

Figure 2 compares the theoretically predicted unsteadiness factors with experimental results from Nomura et al. [6]. It shows that the unsteadiness factor tends to increase with both temperature and pressure because they both increases the droplet temperature. The theoretical prediction is consistent with the widely recognized physical plausibility of the droplet vaporization at high temperature and pressure conditions. The experimental results scatter around the theoretical prediction, despite discernible discrepancy at some specific conditions.

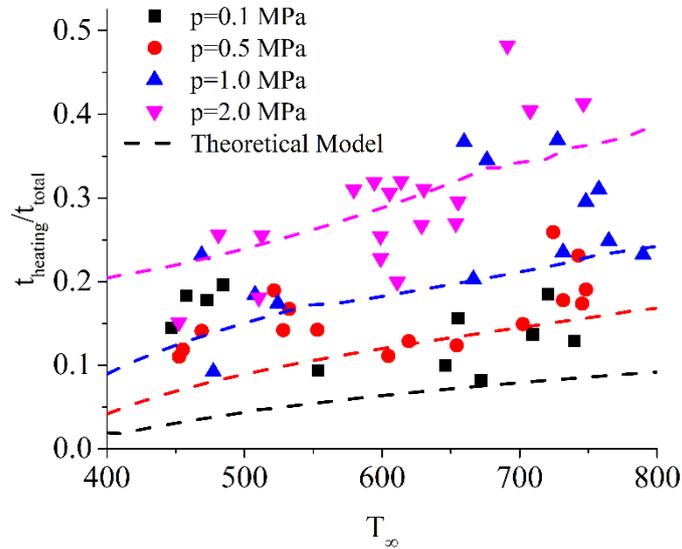

Figure 2. The comparison of theoretically predicted unsteadiness factors with those determined by experimental data [6].

In terms of the droplet lifetime and the unsteadiness factor, we can constitute a theoretical model interpreting the history of droplet surface during the vaporization process. For mathematical convenience, we introduce the normalized droplet diameter and lifetime, defined by

$$d_n = \frac{d}{d_0}, \qquad t_n = \frac{t}{t_{\text{total}}} \tag{67}$$

The droplet size almost remains at the initial instant of the heating stage, i.e.,

$$d_n^2 = 1, \qquad t_n \ll \Phi_i \tag{68}$$

where the unsteadiness factor $\Phi_i$ can be equivalently regarded as the normalized heating time. At the quasi-steady vaporization stage, the droplet surface decays linearly with time, which is consistent with the classic $d^2$-law, i.e.,

$$d_n^2 = 1 - \frac{1}{1-\Phi_i}(t_n - \Phi_i), \qquad t_n > \Phi_i \tag{69}$$

Equations (68) and (69) characterize the asymptotic behaviors of the droplet vaporization process. Therefore, an appropriate vaporization model shall satisfy the subsequent conditions, i.e.,

(I) It must spontaneously become Eqs. (68) and (69) respectively at the initial instant of the heating stage and when the droplet undergoes quasi-steady vaporization.

(II) It must experience a smooth transition from Eq. (68) and (69) at some intermediate instant, which depends upon the unsteadiness factor $\Phi_i$.

To constitute the desired vaporization model, capable to interpret the droplet size history during the whole vaporization process, we define a transition function as

$$S(t) = \frac{\tanh\left[\frac{1}{2\delta_w}\left(\frac{t_n}{\Phi_i} - 1\right)\right] + \tanh\frac{1}{2\delta_w}}{\tanh\left[\frac{1}{2\delta_w}\left(\frac{1}{\Phi_i} - 1\right)\right] + \tanh\frac{1}{2\delta_w}} \tag{70}$$

Equation (70) indicates that $S(t)$ is a normalized function that equals to 0 and 1 at $t_n = 0$ and $t_n = 1$, respectively. The factor $\delta_w$ interprets the width of the transition regime such that within an interval, centered at $\Phi_i$ with width $\delta_w$, the value of the function $S(t)$ changes from 0.1 to 0.9. Thus, the transition function can characterize the smooth transition from heating stage to quasi-steady vaporization stage.

Using the transition function given by Eq. (70), we constitute the theoretical model for droplet vaporization:

$$\frac{d^2}{d_0^2} = 1 - \frac{S(t)}{1-\Phi_i}[t_n - \Phi_i S(t)] \tag{71}$$

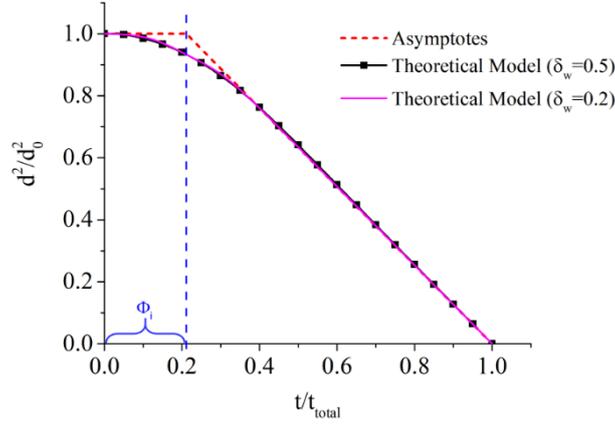

Figure 3. Schematic of theoretical model for droplet vaporization.

Figure 3 shows that the present model for droplet vaporization satisfies both requirements (I) and (II), which verifies its mathematical appropriateness and physical plausibility. Moreover, it indicates that the theoretical model is insensitive to the variation of transition function because the predicted droplet size history changes negligibly as the transition width changes from 0.2 to 0.5. Thereby, we shall adopt that $\delta_w = 0.5$ in the subsequent calculation based on the vaporization model.

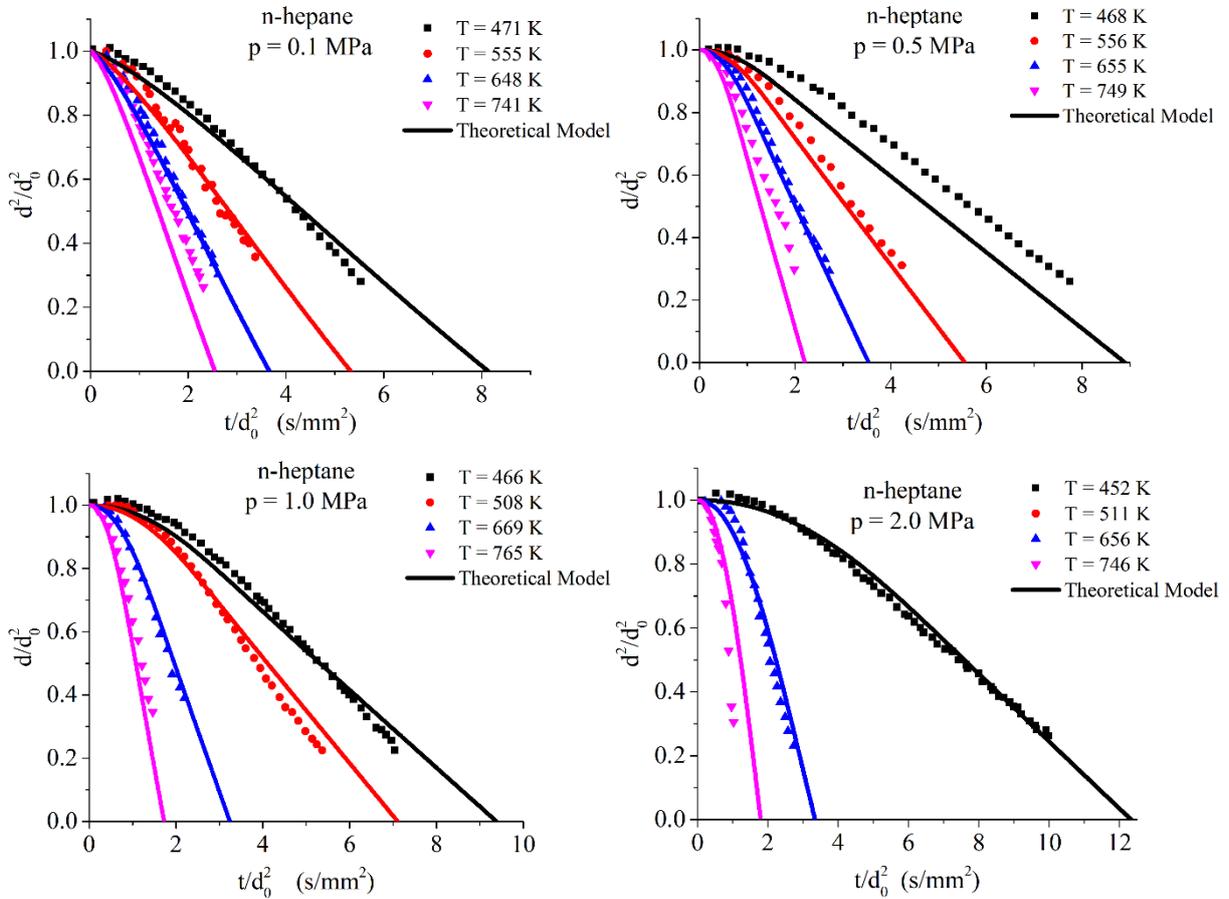

Figure 4. Comparison between the model prediction and experimental data from Nomura et al. [6]

Figure 4 compares the droplet size histories predicted by the theoretical model in Eq. (71) with experimental results reported by Nomura et al. [6]. Good agreement is achieved except for the case of p=0.5 MPa and T=468 K. Overall, the theoretical model is verified by the experiments. As shown in Fig 1, Nomura et al.'s experiments indicate that the pressure has dual effect upon the vaporization lifetime of the droplet, which relies on the ambient temperature. Specifically, when the ambient temperature is relatively low, e.g., $T_\infty < 500$ K, the droplet lifetime time increases with the pressure. However, for $T_\infty > 650$ K, the droplet lifetime decreases with the pressure. Figure 5 shows the variation of theoretically predicted droplet lifetimes with ambient pressure at selected temperatures, indicating that the present theoretical model can consistently reveal the coupled effects of pressure and temperature upon the droplet lifetime. Besides, its intrinsic cause can be interpreted by deeper analysis towards the explicit formulas for heating and vaporization times, given by Eqs. (52) and (55), respectively.

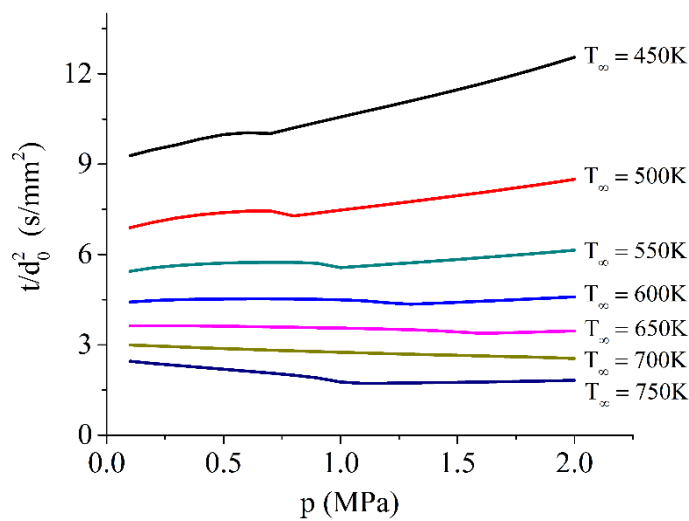

Figure 5. Coupled effect of pressure and temperature upon the droplet lifetime.

According to Fig. 2 the unsteadiness factor $\Phi_i$ grows as the pressure increases. This partially increases the droplet lifetime. Meanwhile, at relatively low temperatures, e.g., $T_\infty < 550$ K, the temperature discrepancy between the ambience and droplet surface would be moderate, which allows us to assume that $c_{pg}(T_\infty - T_s^0) \ll L(T_s^0)$. Consequently, the vaporization time can be simplified to

$$t_{\text{vaporization}} \approx \frac{1}{2} \frac{r_0^2 \rho_l L(T_s^0)}{\lambda_g (T_\infty - T_s^0)} \tag{72}$$

Besides, the surface temperature $T_s^0$ must be remote from the critical temperature, leading to that according to ref [32], the enthalpy of vaporization could be represented by

$$\frac{L(T_s^0)}{L_0} \approx 1 - f_E \left(\frac{q_{l,R}}{2} - 1\right) \frac{T_s^0 - T_R}{T_c - T_R} \tag{73}$$

where the energy factor $f_E$, for the current n-heptane fuel is around 0.055. Taking derivative of $t_{\text{vaporization}}$ with respect to $T_s^0$ from Eq. (72) with simplified enthalpy of vaporization given by Eq. (73), we have

$$\frac{dt_{\text{vaporization}}}{dT_s^0} \approx \frac{1}{2} \frac{r_0^2 \rho_l}{\lambda_g} \frac{L_0}{(T_\infty - T_s^0)^2} \left[1 - f_E \left(\frac{q_{l,R}}{2} - 1\right)\right] > 0 \tag{74}$$

where the reference temperature is 300 K. Therefore, at relatively low temperatures, the vaporization time also increases with the droplet surface temperature, which becomes higher as ambient pressure grows. This is because the heat transfer rate from the ambience decreases as the droplet surface temperature increasing, which lowers the temperature difference between the droplet surface and the ambience. At relatively low temperatures, the decrement of vaporization enthalpy due to the same cause tends to be immaterial so that the vaporization rate is reduced. Thereby, at relatively low pressures, in addition to the lengthening of droplet heating time, the droplet lifetime is noticeably prolonged with increasing pressure. At moderate pressures, Eq. (73) may no longer hold, and consequently the drop of vaporization enthalpy due to increasing of droplet surface temperature renders the phase transition to occur more readily hence accelerates the vaporization process. The reduction of vaporization time tends to compensate the lengthening of the heating time, which leads to the slightly non-monotonicity for droplet lifetimes at intermediate pressures as indicated in Figure 5. At relatively high pressures, the duration of heating stage becomes comparable with the droplet lifetime. The lengthening of heating time due to pressure rise inclines to dominate over its facilitating effect on vaporization rate, which is limited by the environmental temperature, thereby, it leads to monotonic increasing of droplet lifetime with pressure again.

At intermediate temperatures, say 550K $< T_\infty <$ 650K, the facilitating effect of pressure rising (due to lowering the enthalpy of vaporization) and the its impeding effect (due to increasing the heating period) on droplet lifetime tends to balance. Thereby, it results in indiscernible change of droplet lifetime at wide range of pressures.

At high temperatures, e.g., $T_\infty > 650K$, the logarithmic function in Eq. (55) cannot be estimated by its first order Taylor series. However, the dependence of vaporization time upon pressure, i.e., the behavior of $dt_{\text{vaporization}}/dT_s^0$, can still be interpreted by means of Eq. (72) because the logarithmic function has the same monotonicity as linear function. When the droplet surface temperature is close to the critical temperature, according to Ref. [32], the enthalpy of vaporization can be evaluated through

$$\frac{L(T_s^0)}{L_0} \approx f_E \frac{q_{l,R}}{2(1-\alpha)} \left(\frac{T_c - T_s^0}{T_c - T_R}\right)^{1-\alpha} + \left(1 - \frac{1}{2} f_E q_{l,R}\right) \left(\frac{T_c - T_s^0}{T_c - T_R}\right)^\beta \tag{75}$$

Taking derivative of $t_{\text{vaporization}}$ with respect to $T_s^0$ from Eq. (72) with enthalpy of vaporization given by Eq. (75), we have

$$\begin{aligned}\frac{dt_{\text{vaporization}}}{dT_s^0} &\approx \frac{1}{2} \frac{r_0^2 \rho_l}{\lambda_g} \frac{L_0}{(T_\infty - T_s^0)^2} \left[\frac{1}{2} f_E q_{l,R} \left(\frac{T_c - T_s^0}{T_c - T_R}\right)^{1-\alpha} \left(\frac{1}{1-\alpha} - \frac{T_\infty - T_s^0}{T_c - T_s^0}\right)\right.\\ &\left. + \left(1 - \frac{1}{2} f_E q_{l,R}\right) \left(\frac{T_c - T_s^0}{T_c - T_R}\right)^\beta \left(1 - \beta \frac{T_\infty - T_s^0}{T_c - T_s^0}\right)\right]\end{aligned} \tag{76}$$

When the environmental temperature is considerably higher than the critical temperature, $T_s^0$ becomes comparable with $T_c$. At conditions that $(T_\infty - T_s^0)/(T_c - T_s^0) > 1/\beta$, Eq. (76) indicates that $dt_{\text{vaporization}}/dT_s^0$ is negative and thereby the vaporization rate increases with the ambient pressure. This effect can be attributed to that the enthalpy of vaporization becomes exceedingly low when the droplet surface temperature approaching to the critical temperature and thereby facilitating the vaporization process. The reduction of $t_{\text{vaporization}}$ due to the exceedingly decay of enthalpy of vaporization would enforce $t_{\text{heating}}$ to undertake a similar change because according to Fig 2, the unsteadiness factors are below 0.5, implying that $t_{\text{heating}} < t_{\text{vaporization}}$ for all concerned situations. Therefore, at high temperatures, the droplet lifetime decays with the environmental pressure. Interestingly, at exceedingly high temperatures, e.g., $T_\infty > 750K$, the theoretically predicted droplet lifetime tends to increase again with pressure. It might be attributed to that the droplet vaporization process becomes highly unsteady so that the heating time is considerably longer than the quasi-steady vaporization time, which leads to the droplet lifetime increase with pressure again.

## 4. Conclusions

We analyze the droplet vaporization problem based on fundamental principles of fluid mechanics and thermodynamics. The transient governing equations are analytically solved after appropriate coordinate transformations. In combination of the matching conditions and the Clausius-Clapeyron relation, the mass fraction of fuel vapor and temperature on the droplet surface can be iteratively determined by Eqs. (45) and (57). Then the characteristic times for droplet heating and vaporization are respectively calculated through Eqs. (52) and (55), whose sum yields the droplet lifetime for vaporization. The theoretically predicted lifetimes agree well with those extrapolated from experiments by Nomura et al. [6]. In terms of unsteadiness factor, defined by Eq. (66), and the transition function, defined by Eq. (70), a theoretical model is constituted, which illustrates the time change of droplet size in the entire vaporization process, given by Eq. (71). It is verified through comparison with the experimental results within wide ranges of temperature (from 450 K to 750 K) and pressure (from 0.1 MPa to 2 MPa). Besides, the theoretical model successfully reveals the conventional recognition that the droplet diameter approximately remains constant at the heating stage, while during the quasi-steady vaporization, the droplet surface shrinks linearly with time according to the $d^2$-law.

Both experimental observation and the theoretical prediction indicates that the pressure has a dual effect upon droplet lifetime. This is pertinently elucidated based on our theoretical analysis with the help of explicit formula for enthalpy of vaporization, given by Eq. (59). At relative low temperatures below 500K, the droplet lifetime increases with pressure since the heat transfer from the hot ambience to the droplet surface decreases. On the other hand, at relative high temperatures above 700K, increasing the ambient pressure tends to shorten the droplet lifetime. This is because the enthalpy of vaporization decays as the droplet temperature approaches the critical point, which facilitates the vaporization process. At moderate temperatures, the droplet lifetime tends to be independent of the ambient pressure, which would be attributed to the balance between the preceding facilitating and impeding effects of pressure rise upon droplet lifetime.

## Acknowledgement

This work was supported by National Natural Science Foundation of China (Nos. 91841302 and

91741126).